\documentclass{desyproc}
\usepackage{wrapfig}

\begin{document}
\title{High and low mass Axion Haloscopes at UWA}

\author{{\slshape{Ben T. McAllister, Stephen R. Parker, Eugene N. Ivanov, and Michael E. Tobar}}\\
ARC Centre of Excellence for Engineered Quantum Systems, School of Physics, University of Western Australia, 35 Stirling Highway, Crawley 6009, WA, Australia.}

\contribID{mcallister_ben}

\confID{13889}  
\desyproc{DESY-PROC-2016-XX}
\acronym{Patras 2016} 
\doi  

\maketitle

\begin{abstract}
We consider the design of a haloscope experiment (ORGAN) to probe for axions at 26.6 GHz. The motivation for this search is to perform the first direct test of a result which claims a possible axion signal at this frequency. There are many technical issues and optimisations that must be considered in the design of a high mass axion haloscope. We discuss the current status of the ORGAN experiment, as well as its future. We also discuss low mass axion haloscopes employing lumped 3D LC resonators.
\end{abstract}

\section{Introduction}
Axions are a well known solution to the strong CP problem in QCD, proposed by Peccei and Quinn in 1977~\cite{axion}. In addition to providing an elegant solution to the strong CP problem, the axion is a compelling dark matter candidate~\cite{cdm}. Many searches for axions are currently underway, employing a variety of detection techniques. One such technique, known as a haloscope, exploits the axion-two photon coupling, in a process known as the inverse Primakoff effect. For an in depth discussion of haloscopes see~\cite{haloscope1,haloscope2}. The Frequency and Quantum Metrology group at the University of Western Australia is constructing a haloscope to scan for high mass axions, with corresponding photon frequencies around 26 GHz. One motivation for this search is to perform the first direct test of a claimed potential axion signal~\cite{beck1}. This work suggests that axions entering the weak link region of Josephson junctions be responsible for the anomalous Shapiro step-like features seen in a number of experiments. It is claimed that axions with a mass of roughly 110~$\mu$eV could create such an effect. This result is surprising, and has been considered potentially spurious, however, as the axion resides in a large and almost unbounded parameter space any candidate signals merit further investigation. Designing a haloscope at high mass presents many difficulties, which can be understood when one considers the expected power in a resonant cavity due to axion conversion (eq. 1)~\cite{haloscope1}.
\begin{equation}
\centering
P_a=g_{a\gamma\gamma}^2~V~B^2~\frac{\rho_a}{m_a}~C~Q
\end{equation}
Where V is the detecting cavity volume, B is the strength of the magnetic field, $\rho_a$ and $m_a$ are the local axion density and axion mass respectively, Q is the loaded cavity quality factor (provided it is less than the expected axion quality factor of $\sim10^6$) and C is a mode dependent form factor. Firstly, the resonant frequencies of cavities are inversely proportional to radius, and thus the detector volume decreases with higher frequencies. Furthermore, the quantum noise limit and surface resistivity of metals increase at higher frequencies, thus increasing the level of background noise from amplifiers, and decreasing the mode quality factor. Our experiment, ORGAN, is nearly ready to commence its initial pathfinder experiment, and we have recently secured ARC funding through the Centre of Excellence for Engineered Quantum Systems, which will enable us to continue this work.
\begin{figure}[t]
	\centering
	\includegraphics[width=0.4\textwidth]{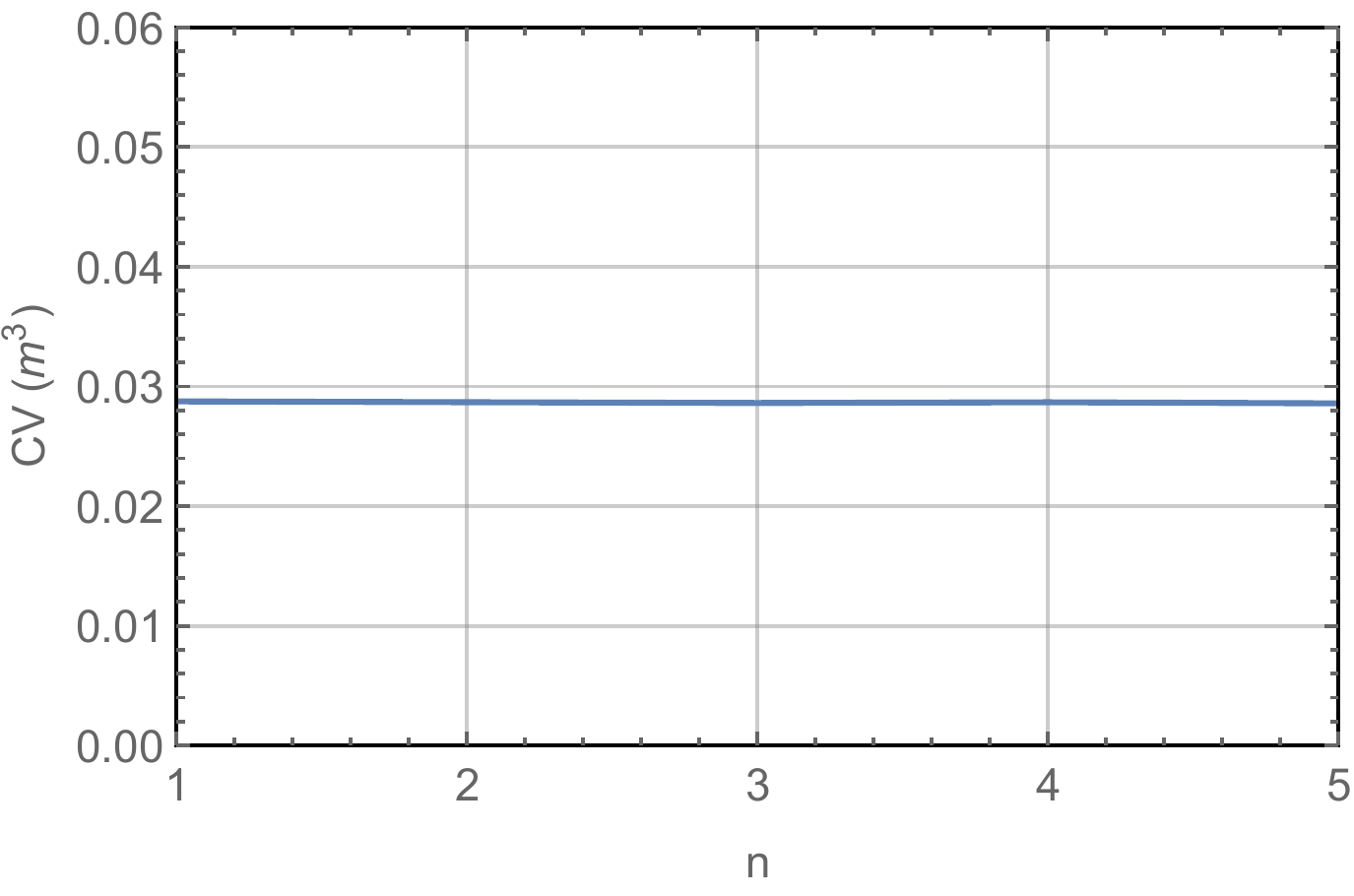}
	\caption{The product of form factor and volume for $TM_{0n0}$ modes at a frequency of 1 GHz, for a cavity length of 1~m.}
	\label{fig:CV}
\end{figure}
\section{ORGAN: status and future}
\begin{figure}[t]
\centering
	\includegraphics[width=0.43\textwidth]{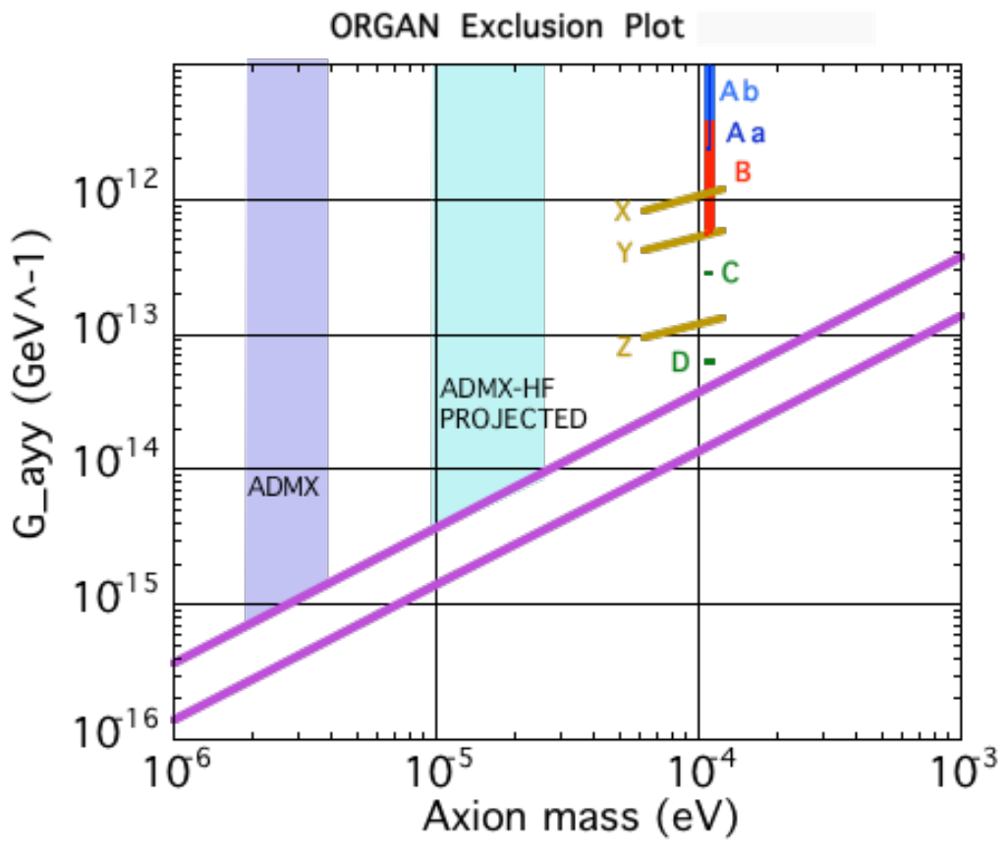}
	\caption{Projected exclusion limits for ORGAN. Aa and Ab represent searches with the current equipment, utilizing 2 cavities. B, C and D represent later phases. Changes include moving from 2 to 8 cavities, implementing a quantum limited JPA in place of the existing amplifier, upgrading the magnet from 7 T to 14 T and improving the quality factor of the cavity using lower loss materials. X, Y and Z represent larger searches at high frequency, with similar experimental scale-ups.}
	\label{fig:Ex}
\end{figure}
\begin{figure}[t]
	\centering
	\includegraphics[width=0.43\textwidth]{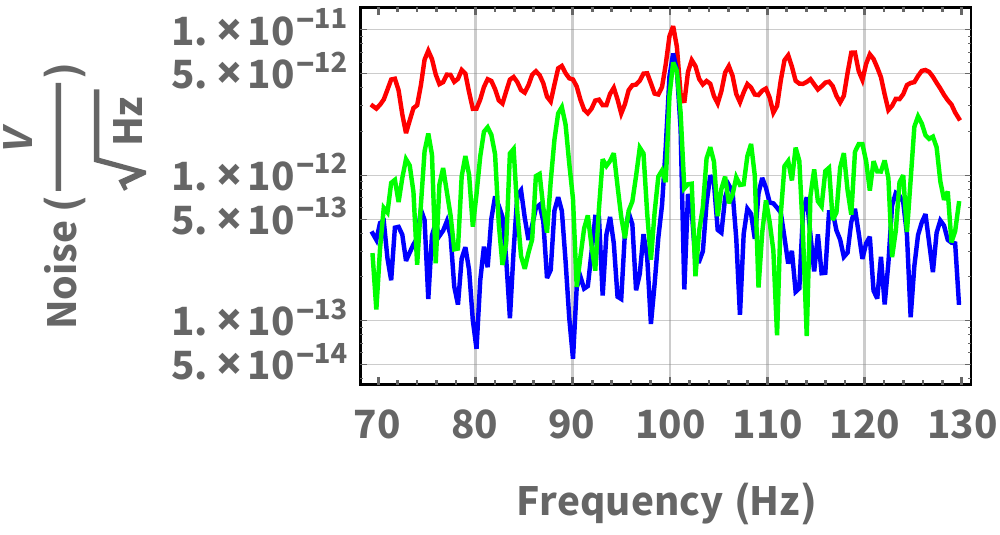}
	\caption{Data from a proof of concept experiment. The PSD is the power spectral density for a single channel with a simulated axion signal injected, the XPSD is the cross-power spectral density for two of the channels, whilst the AXPSD is the averaged cross-power spectral density for the 6 independent cross spectra computed from 4 separate channels.}
	\label{fig:AXPSD}
\end{figure}
The Cry{\bfseries{O}}genic {\bfseries{R}}esonant {\bfseries{G}}roup {\bfseries{A}}xion Co{\bfseries{N}}verter is a haloscope experiment with the goal of synchronizing and combining many resonant cavities, in order to achieve sensitive axion searches at high frequencies, 20 GHz and beyond. The pathfinder experiment will explore a narrow region around 26.6 GHz. For this initial run, a single copper resonant cavity will be employed. This cavity has a radius of $\sim$1 cm, and a length of 5 cm. The $TM_{020}$ mode frequency of this cavity is ~26.55 GHz at 20 mK. A radially moving metallic rod will provide $\sim$1 GHz of tuning, which is enough to cover the entire range of the Beck result~\cite{beck2}. Commonly the $TM_{010}$ mode is chosen for axion haloscopes, as the form factor, $\text{C}_{010}$, is highest for this mode. Whilst the form factor decreases for higher order TM modes, the corresponding cavity radius at a given frequency increases, which leads to an increase in volume. For a given cavity length and a given desired mode frequency the product of C and V can be shown to be essentially constant, as displayed in fig.~\ref{fig:CV}. The result of this is that there is no overall decrease in sensitivity associated with the form factor decrease in a higher order TM mode. Furthermore we have found, and it has been discussed elsewhere~\cite{kinion}, that cavities with lengths greater than $\sim$5 times their radius are disadvantageous to use in haloscope searches, as many length dependent modes crowd around the detector mode, making mode identification and tuning difficult. When this is considered it may be advantageous to employ higher order modes, as the larger radius means a larger allowable length, which means a larger C~V product. The limitation to this approach is the space within the magnet. After the pathfinder run, the experiment will undergo several phases. Figure~\ref{fig:Ex} outlines the projected sensitivity, and details the experimental parameters of the different phases. Many proposals for high frequency haloscopes require synchronisation of multiple cavities~\cite{multicav}, this provides an engineering challenge in practice. We have been exploring novel techniques to achieve this kind of synchronisation, whilst simultaneously improving the signal to noise ratio of axion signals. Cross-correlation measurements are two channel measurement schemes where the cross-spectrum is computed, rejecting uncorrelated noise sources while still retaining correlated signals such as those generated in multiple cavities by axion conversion. Cross-correlating signals from multiple cavities, each with their own amplification chain, presents a method of effectively power combining multiple cavities which can be spatially well separated. It is important to note that this technique alone does not provide an improvement over traditional power summation methods. However, for large numbers of resonators we can achieve an improvement. For n cavities there are $\frac{n(n-1)}{2}$ independent cross-spectra. In each of these cross spectra, the mean of the background noise level is lower, the resolution of a candidate signal is higher, and the standard deviation of the background noise is also higher, when compared with a single channel. If we define SNR as the difference between the peak of the signal and the mean of the background noise, divided by the standard deviation of the background noise, a cross-spectrum provides a factor of 2 improvement in SNR, which is the same as a Wilkinson power combiner. However, if we take these $\frac{n(n-1)}{2}$ independent cross-spectra and average them, we can achieve a factor $2\sqrt{\frac{n(n-1)}{2}}$ improvement in SNR, as this averaging process decreases the standard deviation of the background noise. This effect is shown in fig.~\ref{fig:AXPSD}. For a further discussion of this see~\cite{Xcorr} (update forthcoming). Further research into, and incorporation of this technique is a long term goal of the ORGAN experiment.
\section{New experiments: low mass axion detection}
Our group is developing proposals for new haloscope experiments. One such recent proposal~\cite{RCHalo} considers lumped 3D LC resonators. These resonators, commonly known as re-entrant cavities~\cite{RRcav} consist of metallic cavities containing a central post or ring, and a small gap between the top of this post or ring and the top of the cavity. Adjusting the size of this gap tunes the resonant frequency of the cavity over a large frequency range, which makes these structure very appealing for axion searches. These structures are readily implementable inside existing haloscope infrastructure, and can reach promising low mass axion regimes. Our proposal~\cite{RCHalo} discusses the optimisation of such an experiment, and presents possible axion exclusion limits should such a search be undertaken.
\section{Conclusion}
The FQM group at UWA is developing high and low mass haloscopes. The ORGAN experiment will soon commence its path-finding run, which consists of a single copper resonant cavity employing a traditional low-noise amplifier and a $TM_{020}$ mode for detection. The early stages of the ORGAN experiment will focus on directly testing the claimed Beck result, followed by a wider scan at high frequency. As a part of the future of this experiment we are developing cross-correlation techniques to increase sensitivity. Furthermore, we are pursuing research into new, novel haloscope designs, with the aim of developing systems to sensitively search for axions over many different regions of the parameter space.
\begin{footnotesize}

\end{footnotesize}


\end{document}